\documentclass[11pt]{article}
\pdfoutput=1
\usepackage{jheppub}
\usepackage{amsmath,amssymb,mathtools,bm,youngtab}
\usepackage{enumitem}
\usepackage{xcolor}

\newcommand{\m}{\mathbf m}
\newcommand{\n}{\mathbf n}

\newcommand{\cH}{\mathcal H}
\newcommand{\Sym}{\operatorname{Sym}}
\newcommand{\supp}{\operatorname{supp}}
\newcommand{\cC}{\mathcal C}
\newcommand{\RR}{\mathbb R}
\newcommand{\dd}{\mathrm d}

\begin{document}
\title{Trace Relations and Partition Functions}
\author[a]{Robert de Mello Koch,}
\author[b]{Antal Jevicki,}
\author[b]{and Junjie Zheng}
\affiliation[a]{School of Science, Huzhou University, Huzhou 313000, China}
\affiliation[a]{Mandelstam Institute for Theoretical Physics, School of Physics, University of the Witwatersrand, Private Bag 3, Wits 2050, South Africa}
\affiliation[b]{Department of Physics, Brown University,
182 Hope Street, Providence, RI 02912, United States}
\affiliation[b]{Brown Theoretical Physics Center, Brown University,
340 Brook Street, Providence, RI 02912, United States}
\emailAdd{robert@zjhu.edu.cn}
\emailAdd{antal\_jevicki@brown.edu}
\emailAdd{junjie\_zheng@alumni.brown.edu}
\abstract{In general  trace relations appear in finite $N$ matrix models~\cite{deMelloKoch:2025ngs} and collective field theories. They lead, as was established in~\cite{deMelloKoch:2025rkw}, to reductions of large $N$ Hilbert spaces. Even more importantly they induce the appearance of a whole class of new (miraculous) states whose number grows exponentially in $N^2$~\cite{deMelloKoch:2025ngs}. In this work we discuss their relevance to finite $N$ thermodynamics and the role they play in constraining the partition functions of finite N theories.}
\maketitle

\section{Introduction}

Generally, $U(N)$ matrix theories, $O(N)$ vector field theories and various $S_N$ models all exhibit the phenomenon that we are to discuss. We will therefore consider it in a featured quantum model which has the needed complexity and structure. In finite $N$ matrix models (and consequently vector theories also), trace relations follow from the Cayley-Hamilton theorem applicable to words (invariants) made with $N\times N$ matrices. In $S_N$ quantum mechanical models one has a similar story, while for $S_N$ orbifold CFT's there is a certain confusion. The single trace (primary) invariants and their correlation functions are well understood in the original work~\cite{Jevicki:1998bm}. The single trace invariant operators are defined through $S_N$ averaging. As such they definitely obey finite $N$ trace relations. We will feature them since they have not appeared in the literature. Primaries correspond to center of mass degrees of freedom in the orbifold CFT. The quantum mechanical model governing them will be simply referred to as $S_N$ orbifold.

In some of our previous studies of finite-$N$ invariant algebras \cite{deMelloKoch:2025ngs,deMelloKoch:2025rkw}, the organization of the Hilbert space was particularly transparent because the relevant invariant ring admitted a Hironaka decomposition.  In such cases the Hilbert series takes the form
\begin{equation}
        Z=\frac{P}{Q},
\end{equation}
with the denominator encoding a freely generated set of primary invariants and the all positive numerator encoding the finite set of secondary invariants.  This structure, however, is not generic.  In particular, the $S_N$ model considered here is naturally organized as a sum over different sectors, and its full partition function does not arise from a single Hironaka decomposition. The finite-$N$ trace relations and the Hironaka decomposition remain available in each sector. The present case is of more general nature, where the partition function takes the form
\begin{equation}
Z=\sum_s Z_s\qquad Z_s=\frac{P_s}{Q_s}
\end{equation}
with the sum being over superselection sectors. These sectors are labelled by partitions of $N$, and it is precisely this partition structure that leads to the grand-canonical Hecke form of the partition function. A sectorwise decomposition has also appeared recently in the context of supersymmetric cohomology~\cite{Abhijit}.

The main proposition of this work is that imposition of trace relations constrains the finite $N$ partition function. For the global $S_N$ model that we feature, this leads to the polynomial structure, and grand canonical Hecke form  which is the correct answer for this class of theories. The proposition will be demonstrated, through explicit construction of  trace relation for increasing $N$, and of the corresponding exclusions that they imply on the space of states contributing to partition functions. What becomes visible (and is the main result of the work) is that the exclusions are more complex, and very different from the naively used ``exclusion principle". They are fully encoded in the nonlinear structure of the trace relations.

\section{Trace Relations}
\label{sec:trace_relations}

The $S_N$ primary operators were constructed in detail for a CFT on $K3$ or $T^4$ in~\cite{Jevicki:1998bm}. They take the form
\begin{equation}
O^{(p,\bar p)}=\frac1{\sqrt{N!(N-n)!n!}}\sum_{h\in S_N}F^{(p,\bar p)}O_{h\gamma_n h^{-1}}\label{collectivetrace}
\end{equation}
where $\gamma_n=(1\,2\,\cdots n)$ is the defining (cyclic) permutation and $h$ a general element of $S_N$. Based on this \cite{Jevicki:1998bm} one evaluates  three-point correlators of chiral primaries in the operator scheme. In this work we will consider the \emph{global structure} of these operators, pertinent to trace relations. The only property being used is \emph{factorization} $\sigma(1,2,\cdots,n)=\sigma(1,2)\sigma(2,3)\cdots\sigma(n-1,n)$ in the sense of a matrix, and that the product of two $\sigma$s gives a $\sigma$. This is also the crux of a simple holographic model, introduced in\cite{Jevicki:1997nr}. The set of all single trace operators generally translates into  collective variables
\begin{equation}
O\,\,\longleftrightarrow\,\,\Phi
\end{equation}
which represents an overcomplete set. For notational simplicity  we will only deal with the bosonic fields (coordinates) $\{x_i\}$. Fermions can be similarly included.

In the general expression above for each reference n-cycle $ \gamma_n=(1\,2\,\cdots n)$ there are $(n-1)!$ different $n$-cycles on any fixed $I\subset[N]$. Since cyclic ordering is not counted as an independent collective trace we will work with a \emph{support-normalized} cycle sum
\begin{equation}
\frac{1}{(n-1)!}\sum_{\gamma\in\cC_n}F(\supp\gamma)=\sum_{\substack{I\subset[N]\\|I|=n}}F(I),
\label{eq:support-normalized}
\end{equation}
where $\cC_n$ denotes the conjugacy class consisting of one $n$-cycle and $N-n$ fixed points.  A finite-$N$ cluster configuration is specified by a set partition
\begin{equation}
 \pi=\{B_1,\ldots,B_s\}
\end{equation}
of $[N]=\{1,\ldots,N\}$ together with a center coordinate $X_B\in\RR^d$ for each block $B\in\pi$.  Define the incidence indicator
\begin{equation}
 \chi_I(\pi)=
 \begin{cases}
 1,& I\subset B\text{ for some }B\in\pi,\\
 0,& \text{otherwise}.
 \end{cases}
 \label{eq:chi-def}
\end{equation}
This is a discrete idempotent $\chi_I^2=\chi_I$. If two overlapping subsets are each contained in a block of a set partition, the shared label forces the two blocks to be the same block, i.e. whenever $I\cap J\ne\varnothing$ we have the product
\begin{equation}
 \chi_I\chi_J=\chi_{I\cup J}. \label{eq:union-rule}
\end{equation}
For a given partition configuration $(\pi,\{\vec{X}_B\})$, define
\begin{equation}
\Phi_n(\vec k)=\sum_{\substack{I\subset[N]\\|I|=n}}\chi_I(\pi)\,e^{i \vec{k}\cdot \vec{X}_{B(I)}}.\label{eq:Phi-subset}
\end{equation}
Here $B(I)$ is the unique block containing $I$ when $\chi_I=1$. $\Phi_n$ counts all $n$-element supports inside the cluster. An $r$-cluster contributes $\binom{r}{n}$ times to $\Phi_n$. The particle-number constraint takes the form
\begin{equation}
 \int\dd^dx\,\Phi_1(x)=N, \label{eq:Phi1-N}
\end{equation}
because an exact block of size $r$ contributes $r$ to $\Phi_1$. We can also consider the $r$ cluster density $\rho_r(\vec{k})$. The translation between $\Phi_n$ and the cluster densities is
\begin{equation}
 \Phi_n=\sum_{r=n}^{N}\binom{r}{n}\rho_r. \label{eq:zeta-forward}
\end{equation}
The matrix $A_{nr}=\binom{r}{n}$ is triangular with unit diagonal, and can thus be inverted to give
\begin{equation}
\rho_r=\sum_{n=r}^{N}(-1)^{n-r}\binom{n}{r}\Phi_n.\label{eq:zeta-inverse}
\end{equation}
For $N=3$, for example
\begin{align}
\Phi_1&=\rho_1+2\rho_2+3\rho_3,\qquad\Phi_2=\rho_2+3\rho_3,\qquad \Phi_3=\rho_3, \label{eq:N3-zeta}
\end{align}
and hence
\begin{align}
\rho_3&=\Phi_3,\qquad \rho_2=\Phi_2-3\Phi_3,\qquad \rho_1=\Phi_1-2\Phi_2+3\Phi_3.
 \label{eq:N3-inverse}
\end{align}
The trace relations can be expressed in terms of either field variable $\Phi_n$ or $\rho_n$.

Generally trace relations (in Fock space) determine the structure of the Hilbert space. The main thrust of the present work is to show that they also fully determine the structure 
of finite-$N$ partition functions.

We will use a ``momentum-space'' notation. Let ${\bf m}=(m_1,m_2)$ be oscillator quantum numbers and write
\begin{equation}
A_n({\bf m})=\sum_{\substack{I\subset[N]\\ |I|=n}}a_I({\bf m}).
\end{equation}
For a fixed microscopic cluster $I$, multiplication adds the mode labels,
\begin{equation}
a_I({\bf m})a_I({\bf n})=a_I({\bf m}+{\bf n}).
\end{equation}
If $I_1,\ldots,I_r$ have connected overlap graph, all particles in $I_1\cup\cdots\cup I_r$ coincide and
\begin{equation}
\prod_{\alpha=1}^{r}a_{I_\alpha}({\bf m}_\alpha)=a_{I_1\cup\cdots\cup I_r}\!\left(\sum_{\alpha=1}^{r}{\bf m}_\alpha\right).
\end{equation}
If the overlap graph has several connected components, each component gives one joined cluster and distinct components remain disjoint.

The crucial finite-$N$ restriction is immediate: no set of mutually disjoint supports can use more than the available $N$ microscopic labels.  It is this simple fact, expressed nonlinearly in the $A_n$, that truncates the large-$N$ collective algebra.

The product of two collective traces separates naturally into overlapping and disjoint contributions.  Consider
\begin{equation}
 A_n({\bf m})A_m({\bf n})
 =\sum_{\substack{|I|=n\\ |J|=m}}a_I({\bf m})a_J({\bf n}).
\end{equation}
Split the sum according to $t=|I\cap J|$.  For $t\geq1$ the two supports join, and their union has size
\begin{equation}
 |I\cup J|=n+m-t.
\end{equation}
For a fixed union $U$ of this size, the number of ordered pairs $(I,J)$ with $|I|=n$, $|J|=m$ and $|I\cap J|=t$ is
\begin{equation}
 C_{nmt}=\frac{(n+m-t)!}{t!\,(n-t)!\,(m-t)!}.
 \label{eq:Cnmt-new}
\end{equation}
Denoting the contribution from disjoint supports by $D_{n,m}({\bf m},{\bf n})$, we therefore obtain the universal product rule
\begin{equation}
 A_n({\bf m})A_m({\bf n})
 =D_{n,m}({\bf m},{\bf n})
 +\sum_{t=1}^{\min(n,m)}C_{nmt}\,
 A_{n+m-t}({\bf m}+{\bf n}) .
 \label{eq:overlap-master}
\end{equation}
As usual $A_r=0$ for $r>N$.  The first term is the genuinely new multi-cluster sector; the remaining terms arise from joining overlapping supports.

For example,
\begin{equation}
 A_1({\bf m})A_1({\bf n})
 =D_{1,1}({\bf m},{\bf n})+A_1({\bf m}+{\bf n}),
 \label{eq:A1A1-example}
\end{equation}
and
\begin{equation}
 A_2({\bf m})A_2({\bf n})
 =D_{2,2}({\bf m},{\bf n})+A_2({\bf m}+{\bf n})+6A_3({\bf m}+{\bf n}).
 \label{eq:A2A2-example}
\end{equation}
At $N=3$, two disjoint pairs cannot exist and hence $D_{2,2}=0$.  At $N=4$, $D_{2,2}$ is nonzero and gives the new $[2,2]$ sector.  This illustrates the general mechanism: increasing $N$ turns previously vanishing disjoint pieces into new independent sectors.

The finite-$N$ relations can be obtained directly by inclusion--exclusion on the microscopic supports.  Start with a product of $r$ singleton traces.  It contains the term in which all $r$ microscopic labels are distinct, together with terms in which two or more labels coincide.  Subtract the pairwise coincidences, restore the terms that were subtracted more than once, and continue in this way until every coincidence pattern has been accounted for.

A compact way to write the result is to let $\Pi_r$ denote the set of ways of grouping $\{1,\ldots,r\}$ into disjoint blocks.  For a grouping $\pi\in\Pi_r$, define
\begin{equation}
 {\bf m}_B=\sum_{i\in B}{\bf m}_i.
\end{equation}
The contribution in which the $r$ singleton supports are mutually disjoint is then
\begin{equation}
 D_{1^r}({\bf m}_1,\ldots,{\bf m}_r)
 =\sum_{\pi\in\Pi_r}
 \left[\prod_{B\in\pi}(-1)^{|B|-1}(|B|-1)!\right]
 \prod_{B\in\pi}A_1({\bf m}_B).
 \label{eq:singleton-inclusion-exclusion}
\end{equation}
The coefficients $1,-1,2,-6,24,\ldots$ arise simply because a block of $k$ coincident labels can be assembled in $(k-1)!$ connected ways, with the alternating sign dictated by inclusion--exclusion.

Equation~\eqref{eq:singleton-inclusion-exclusion} has an immediate finite-$N$ consequence.  Since $N+1$ mutually disjoint microscopic labels do not exist,
\begin{equation}
 D_{1^{N+1}}({\bf m}_1,\ldots,{\bf m}_{N+1})=0.
 \label{eq:fundamental-singleton-trace}
\end{equation}
This is the basic singleton trace relation at finite $N$.  For example, at $N=2$ it gives
\begin{align}
0={}&A_1({\bf m}_1)A_1({\bf m}_2)A_1({\bf m}_3)
-A_1({\bf m}_1+{\bf m}_2)A_1({\bf m}_3)\nonumber\\
&-A_1({\bf m}_1+{\bf m}_3)A_1({\bf m}_2)
-A_1({\bf m}_2+{\bf m}_3)A_1({\bf m}_1)
+2A_1({\bf m}_1+{\bf m}_2+{\bf m}_3),
\end{align}
which is precisely the relation excluding a three-singleton sector.

The same direct subtraction works for mixed clusters.  One first joins any overlapping microscopic supports; the remaining terms are then organized according to which joined clusters are mutually disjoint.  Thus the full hierarchy of mixed trace relations follows from one elementary rule: overlapping supports join, while a collection of disjoint clusters cannot use more than the available $N$ microscopic labels.

The preceding discussion gives a simple interpretation of the physical finite-$N$ Hilbert space.  After all overlaps have been reduced, an independent configuration is specified only by the sizes of its mutually disjoint connected clusters.  Hence a normal form is labelled by an integer partition $\lambda=[1^{\lambda_1}2^{\lambda_2}\cdots]$, with $\sum_{r\geq1}r\lambda_r=N$. The role of the trace relations is therefore to reduce arbitrary products of collective traces to these partition-labelled normal forms.  Equivalently, a putative sector is not needed whenever the total number of microscopic labels required by its mutually disjoint clusters exceeds $N$.This observation is the bridge to the partition function.  Once the one-cluster spectrum is known, the contribution of a partition $\lambda$ is obtained by symmetrizing identical clusters of each size, and the finite-$N$ partition function is the sum over $\lambda\vdash N$.  The low-$N$ examples below make this reduction explicit and then lead directly to the grand-canonical/Hecke form.

We also have the following fact of relevance to trace relations. Based on all traces as variables one can build a collective (field) theory with a cubic Hamiltonian with $G=1/N$ appearing as a coupling constant. This theory has one further, most important property: for integer $N$ the trace relations appear as null states of the Hamiltonian. This can be seen on general grounds, it has also been explicitly demonstrated in a number of cases. One is therefore able to study, generate and confirm trace relations through consistency with the collective field theory. For the above $S_N$ trace relations we have confirmed this property.

\section{Partition Functions from  Trace Relations}
Free energies for $S_N$ orbifolds are well studied in \cite{Bantay:1999us}. Our interest is to show that these follow from the infinite set of nonlinear trace relations at finite-$N$ .
We now apply the support algebra developed above.  The collective traces $A_n({\bf m})$ and the microscopic joining rule have already been defined in Sections \ref{sec:trace_relations}.  At fixed $N$, the trace relations determine which mutually disjoint cluster sectors survive and therefore determine the finite-$N$ partition function.

\paragraph{$N=1$:} There is only one microscopic particle. The trace relation
\begin{equation}
A_1(\m_1)A_1(\m_2)-A_1(\m_1+\m_2)=0.
\end{equation}
says there can't be two distinct singletons. Repeated action on the same microscopic particle merely adds oscillator labels. Also $A_k=0$ for $k>1$.  Hence the only sector is $1=[1]$. Thus
\begin{equation}
Z_1=Z(q).
\end{equation}

\paragraph{$N=2$:} Now two distinct singleton labels do exist.  The disjoint piece is generated by the trace relation of $N=1$
\begin{equation}
A_1(\m_1)A_1(\m_2)-A_1(\m_1+\m_2)=D_{1,1}(\m_1,\m_2)
\end{equation}
which is now non-zero. The trace relations imply that three distinct singleton labels are impossible:
\begin{align}
0={}&A_1(\m_1)A_1(\m_2)A_1(\m_3)-A_1(\m_1+\m_2)A_1(\m_3)-A_1(\m_1+\m_3)A_1(\m_2)\nonumber\\
&-A_1(\m_2+\m_3)A_1(\m_1)+2A_1(\m_1+\m_2+\m_3).
\end{align}
Thus $[1,1]$ is allowed but $[1,1,1]$ (or any higher disjoint sector) is not. There is one microscopic pair $\{1,2\}$, so $A_2$ is nonzero.  We have the trace relation
\begin{equation}
A_1(\n)A_2(\m)=2A_2(\m+\n).\label{12tr}
\end{equation}
Thus a singleton cannot be disjoint from the 2-cluster ($D_{1,2}=0$) and there is no $[2,1]$ sector at $N=2$. Likewise there is only one pair support
\begin{equation}
A_2(\m_1)A_2(\m_2)=A_2(\m_1+\m_2).
\end{equation}
Thus repeated $A_2$ factors do not create an independent $[2,2]$ sector, or any higher disjoint sector; they merely add oscillator labels on the unique two-particle support. Thus, the only contributing sectors are $[1,1]$ and $[2]$. 

The paper~\cite{deMelloKoch:2025eqt} has already demonstrated how the trace relations determine the primary and secondary invariants, which then determine the form of the partition function. For this problem, in the $[1,1]$ sector, the complete set of primary invariants are given by $A_1(1,0)$, $A_1(0,1)$, $A_1(2,0)$ and $A_1(0,2)$, and there is a trivial as well as a single non-trivial secondary invariant given by $A_1(1,1)$. In the $[2]$ sector we have two primary invariants $A_2(1,0)$ and $A_2(0,1)$ and a single trivial secondary invariant. Using these invariants the partition function is determined and the result is a sum of two terms, each of the Hironaka form
\begin{equation}
    Z_2(q)=Z_{[1,1]}+Z_{[2]}=q^{-2}\frac{1+q^2}{(1-q)^2(1-q^2)^2}+\frac{q^{-1/2}}{(1-q^{1/2})^2} .
\end{equation}
This can be shown to be equivalent to
\begin{equation}
Z_2={1\over 2}\left[Z^2(q)+Z(q^2)\right]+Z(\sqrt{q})\label{eq:Z2}
\end{equation}
with a generalized Schur form.

\paragraph{$N=3$:} The trace relation we had for three distinct singletons no longer holds because we can have three distinct particles
\begin{align}
&A_1(\m_1)A_1(\m_2)A_1(\m_3)-A_1(\m_1+\m_2)A_1(\m_3)-A_1(\m_1+\m_3)A_1(\m_2)\nonumber\\
&-A_1(\m_2+\m_3)A_1(\m_1)+2A_1(\m_1+\m_2+\m_3)=D_{1,1,1}(\m_1,\m_2,\m_3)\ne 0.
\end{align}
Thus, the sector $[1,1,1]$ exists. However, there is a trace relation that forbids four distinct particles
\begin{align}
0={}&A_1(\m_1)A_1(\m_2)A_1(\m_3)A_1(\m_4)-A_1(\m_1+\m_2)A_1(\m_3)A_1(\m_4)-A_1(\m_1+\m_3)A_1(\m_2)A_1(\m_4)\nonumber\\
&-A_1(\m_1+\m_4)A_1(\m_2)A_1(\m_3)-A_1(\m_2+\m_3)A_1(\m_1)A_1(\m_4)-A_1(\m_2+\m_4)A_1(\m_1)A_1(\m_3)\nonumber\\
&-A_1(\m_3+\m_4)A_1(\m_1)A_1(\m_2)+2A_1(\m_1+\m_2+\m_3)A_1(\m_4)+2A_1(\m_1+\m_2+\m_4)A_1(\m_3)\nonumber\\
&+2A_1(\m_1+\m_3+\m_4)A_1(\m_2)+2A_1(\m_2+\m_3+\m_4)A_1(\m_1)+A_1(\m_1+\m_2)A_1(\m_3+\m_4)\nonumber\\
&+A_1(\m_1+\m_3)A_1(\m_2+\m_4)+A_1(\m_1+\m_4)A_1(\m_2+\m_3)-6A_1(\m_1+\m_2+\m_3+\m_4)\nonumber\\
={}&D_{1,1,1,1}(\m_1,\m_2,\m_3,\m_4)= 0.
\end{align}
The trace relation \eqref{12tr} is modified to
\begin{equation}
A_1(\n)A_2(\m)-2A_2(\m+\n)=D_{1,2}(\n,\m)\neq 0
\end{equation}
$D_{1,2}$ contains the microscopic terms $a_{12}a_3$, $a_{13}a_2$, $a_{23}a_1$. Thus, we now have a $[2,1]$ sector. A pair plus two disjoint singleton clusters would require four microscopic particles so it can not exist. This is proved using the trace relation
\begin{align}
0={}&A_2(\m)A_1(\n_1)A_1(\n_2)-A_2(\m)A_1(\n_1+\n_2)\nonumber\\
&-2A_2(\m+\n_1)A_1(\n_2)-2A_2(\m+\n_2)A_1(\n_1)+6A_2(\m+\n_1+\n_2).\label{sectr}
\end{align}
The three-particle cluster $A_3$ exists at $N=3$. Two disjoint pairs would require four particles so it should not exist. This is proved by the trace relation
\begin{equation}
A_2(\m)A_2(\n)=A_2(\m+\n)+6A_3(\m+\n).
\end{equation}
A 3-cluster plus a disjoint singleton would require four particles, so it should not exist. This is easily proved by the trace relation
\begin{equation}
A_1(\n)A_3(\m)=3A_3(\m+\n).\label{rerel13}
\end{equation}
Finally, there is only one three cluster because
\begin{equation}
A_3(\m_1)A_3(\m_2)=A_3(\m_1+\m_2).
\end{equation}
Thus, the trace relations prove that the only sectors that contribute are $[1,1,1]$, $[2,1]$ and $[3]$. Again solving the trace relations for a suitable set of primary and secondary invariants we obtain the partition function as a sum over three sectors, with each term taking the Hironaka form
\begin{equation}
    Z_3=Z_{[1,1,1]}+Z_{[2,1]}+Z_{[3]}.
\end{equation}
This can be seen to be identical to the expression
\begin{eqnarray}
Z_3&=&{1\over 6}\left[Z(q)^3+3Z(q)Z(q^2)+2Z(q^3)\right]+Z(\sqrt{q})Z(q)+Z(q^{1\over 3}).\label{eq:Z3}
\end{eqnarray}

\paragraph{$N=4$:} At $N=3$ the quartic singleton expression vanished because four distinct microscopic particles could not exist.  At $N=4$ the trace relation is now nonzero:
\begin{align}
&A_1(\m_1)A_1(\m_2)A_1(\m_3)A_1(\m_4)-A_1(\m_1+\m_2)A_1(\m_3)A_1(\m_4)\nonumber\\
&\quad-A_1(\m_1+\m_3)A_1(\m_2)A_1(\m_4)-A_1(\m_1+\m_4)A_1(\m_2)A_1(\m_3)\nonumber\\
&\quad-A_1(\m_2+\m_3)A_1(\m_1)A_1(\m_4)-A_1(\m_2+\m_4)A_1(\m_1)A_1(\m_3)\nonumber\\
&\quad-A_1(\m_3+\m_4)A_1(\m_1)A_1(\m_2)\nonumber\\
&\quad+2A_1(\m_1+\m_2+\m_3)A_1(\m_4)+2A_1(\m_1+\m_2+\m_4)A_1(\m_3)\nonumber\\
&\quad+2A_1(\m_1+\m_3+\m_4)A_1(\m_2)+2A_1(\m_2+\m_3+\m_4)A_1(\m_1)\nonumber\\
&\quad+A_1(\m_1+\m_2)A_1(\m_3+\m_4)+A_1(\m_1+\m_3)A_1(\m_2+\m_4)\nonumber\\
&\quad+A_1(\m_1+\m_4)A_1(\m_2+\m_3)-6A_1(\m_1+\m_2+\m_3+\m_4)\nonumber\\
&=D_{1,1,1,1}(\m_1,\m_2,\m_3,\m_4)\neq0.\label{eq:D1111}
\end{align}
Thus four mutually disjoint singleton clusters exist. However, five distinct singleton particles cannot exist at $N=4$.  The trace relation which proves this is
\begin{align}
0={}&\prod_{i=1}^{5}A_1(\m_i)-\sum_{1\le i<j\le5}A_1(\m_i+\m_j)\prod_{\ell\neq i,j}A_1(\m_\ell)\nonumber\\
&+2\sum_{1\le i<j<k\le5}A_1(\m_i+\m_j+\m_k)\prod_{\ell\notin\{i,j,k\}}A_1(\m_\ell)\nonumber\\
&+\sum_{\substack{\{\{i,j\},\{k,\ell\}\}\\  \text{two disjoint unordered pairs}}} A_1(\m_i+\m_j)A_1(\m_k+\m_\ell)A_1(\m_s)\nonumber\\
&-6\sum_{i=1}^{5} A_1\!\left(\sum_{j\neq i}\m_j\right)A_1(\m_i)-2\sum_{\substack{S\subset\{1,\ldots,5\}\\ |S|=3}}
 A_1\!\left(\sum_{i\in S}\m_i\right) A_1\!\left(\sum_{j\notin S}\m_j\right)\nonumber\\
&+24A_1(\m_1+\m_2+\m_3+\m_4+\m_5),\label{eq:F5}
\end{align}
where in the third line of equation \eqref{eq:F5}, $s$ denotes the one index not contained in the two displayed pairs.
For equal oscillator labels this is
\begin{align}
0={}&A_1(\m)^5-10A_1(2\m)A_1(\m)^3+15A_1(2\m)^2A_1(\m)+20A_1(3\m)A_1(\m)^2\nonumber\\
&-20A_1(2\m)A_1(3\m)-30A_1(4\m)A_1(\m)+24A_1(5\m).\label{eq:F5equal}
\end{align}
At $N=4$ the trace relation \eqref{sectr} becomes
\begin{align}
&A_2(\m)A_1(\n_1)A_1(\n_2)-A_2(\m)A_1(\n_1+\n_2)\nonumber\\
&\quad-2A_2(\m+\n_1)A_1(\n_2)-2A_2(\m+\n_2)A_1(\n_1)+6A_2(\m+\n_1+\n_2)\nonumber\\
&=D_{2;1,1}(\m;\n_1,\n_2)\neq0.\label{eq:D211}
\end{align}
A microscopic representative is $a_{12}(\m)a_3(\n_1)a_4(\n_2)$. Thus the sector $[2,1,1]$ exists. A pair plus three mutually disjoint singletons requires at least five particles, so it must vanish at $N=4$.  The trace relation that proves this is
\begin{align}
0={}&
A_2(\m)A_1(\n_1)A_1(\n_2)A_1(\n_3)-A_2(\m)\Big[A_1(\n_1+\n_2)A_1(\n_3)+A_1(\n_1+\n_3)A_1(\n_2)
+A_1(\n_2+\n_3)A_1(\n_1)\Big]\nonumber\\
&-2\Big[ A_2(\m+\n_1)A_1(\n_2)A_1(\n_3)+A_2(\m+\n_2)A_1(\n_1)A_1(\n_3)
+A_2(\m+\n_3)A_1(\n_1)A_1(\n_2)\Big]\nonumber\\
&+2\Big[ A_2(\m+\n_1)A_1(\n_2+\n_3)+A_2(\m+\n_2)A_1(\n_1+\n_3)+A_2(\m+\n_3)A_1(\n_1+\n_2)\Big]
\nonumber\\
&+6\Big[ A_2(\m+\n_1+\n_2)A_1(\n_3)+A_2(\m+\n_1+\n_3)A_1(\n_2)+A_2(\m+\n_2+\n_3)A_1(\n_1)\Big]
\nonumber\\
&+2A_2(\m)A_1(\n_1+\n_2+\n_3)-24A_2(\m+\n_1+\n_2+\n_3).\label{eq:D2111}
\end{align}
At $N=4$ we should be able to support a $[2,2]$ sector. The quadratic pair product is
\begin{align}
A_2(\m_1)A_2(\m_2)=A_2(\m_1+\m_2)+6A_3(\m_1+\m_2)+D_{2,2}(\m_1,\m_2).\label{eq:A2A2}
\end{align}
At $N=3$ the disjoint term vanished $D_{2,2}=0$, but at $N=4$ it is nonzero
\begin{align}
A_2(\m_1)A_2(\m_2)-A_2(\m_1+\m_2)-6A_3(\m_1+\m_2)=D_{2,2}(\m_1,\m_2)\neq0.\label{eq:D22}
\end{align}
A microscopic representative is $a_{12}(\m_1)a_{34}(\m_2)$. Thus $[2,2]$ is a genuinely new $N=4$ sector.

A pair-pair configuration together with an additional disjoint singleton would require five microscopic labels. Thus the $[2,2,1]$ sector is excluded at $N=4$. The trace relation that proves this is 
\begin{align}
0={}&A_2(\m_1)A_2(\m_2)A_1(\n)-A_2(\m_1+\m_2)A_1(\n)\nonumber\\
&-6A_3(\m_1+\m_2)A_1(\n)-2A_2(\m_1+\n)A_2(\m_2)\nonumber\\
&-2A_2(\m_1)A_2(\m_2+\n)+4A_2(\mathbf \m_1+\m_2+\n)+24A_3(\mathbf \m_1+\m_2+\n),
\label{eq:A2A2A1}
\end{align}
At $N=3$ the trace relation \eqref{rerel13} showed that a 3-cluster could not have a disjoint singleton. At $N=4$ the equation \eqref{rerel13} becomes
\begin{align}
A_1(\n)A_3(\m)-3A_3(\m+\n)=D_{1,3}(\n,\m)\neq0. \label{eq:D13}
\end{align}
so that a $[3,1]$ sector is allowed. A microscopic representative is $a_4(\n)a_{123}(\m)$. A 3-cluster plus two mutually disjoint singletons would require five particles, so the $[3,1,1]$ sector should not exist. The trace relation that proves this is
\begin{align}
0={}&A_3(\m)A_1(\n_1)A_1(\n_2)-A_3(\m)A_1(\n_1+\n_2)\nonumber\\
&-3A_3(\m+\n_1)A_1(\n_2)-3A_3(\m+\n_2)A_1(\n_1)+12A_3(\m+\n_1+\n_2).\label{eq:D311}
\end{align}
A 3-cluster and a 2-cluster cannot be mutually disjoint at $N=4$.  This is a consequence of the trace relation
\begin{align}
A_2(\m)A_3(\n)=3A_3(\m+\n)+12A_4(\m+\n).\label{eq:A2A3}
\end{align}
The trace relation is
\begin{align}
A_3(\m_1)A_3(\m_2)=A_3(\m_1+\m_2)+12A_4(\m_1+\m_2).\label{eq:A3A3}
\end{align}
The four-particle cluster $A_4$ exists.  Since it already uses all four microscopic particles, no additional cluster can be disjoint from it. The complete set of products involving $A_4$ is
\begin{align}
A_1(\n)A_4(\m)&=4A_4(\m+\n),\\
A_2(\n)A_4(\m)&=6A_4(\m+\n),\\
A_3(\n)A_4(\m)&=4A_4(\m+\n),\\
A_4(\n)A_4(\m)&=A_4(\m+\n).
\label{eq:A4products}
\end{align}

The complete list of quadratic relations are
\begin{align}
A_1(\m)A_1(\n)&=D_{1,1}(\m,\n)+A_1(\m+\n),\\
A_1(\m)A_2(\n)&=D_{1,2}(\m,\n)+2A_2(\m+\n),\\
A_1(\m)A_3(\n)&=D_{1,3}(\m,\n)+3A_3(\m+\n),\\
A_1(\m)A_4(\n)&=4A_4(\m+\n),\\
A_2(\m)A_2(\n)&=D_{2,2}(\m,\n)+A_2(\m+\n)+6A_3(\m+\n),\\
A_2(\m)A_3(\n)&=3A_3(\m+\n)+12A_4(\m+\n),\\
A_2(\m)A_4(\n)&=6A_4(\m+\n),\\
A_3(\m)A_3(\n)&=A_3(\m+\n)+12A_4(\m+\n),\\
A_3(\m)A_4(\n)&=4A_4(\m+\n),\\
A_4(\m)A_4(\n)&=A_4(\m+\n).
\end{align}
The genuinely new disjoint quadratic structures at $N=4$ are $D_{1,3}\neq0$ and $D_{2,2}\neq0$.

Thus the trace relations have established that five sectors $[1,1,1,1]$, $[2,1,1]$, $[2,2],$ $[3,1]$ and $[4]$ contribute. The partition function is therefore a sum over five terms with each of the Hironaka form
\begin{equation}
    Z_4=Z_{[1,1,1,1]}+Z_{[2,2]}+Z_{[3,1]}+Z_{[2,1,1]}+Z_{[4]}
\end{equation}
It can be shown to be equivalent to
\begin{align}
Z_4(q)={}&\frac1{24}\left[Z(q)^4+6Z(q)^2Z(q^2)+3Z(q^2)^2+8Z(q)Z(q^3)+6Z(q^4)\right]\nonumber\\
&+\frac12\left[Z(q)^2+Z(q^2)\right]Z(q^{1/2})+\frac12\left[Z(q^{1/2})^2+Z(q)\right]+Z(q)Z(q^{1/3})+Z(q^{1/4}).\label{eq:Z4}
\end{align}

The formulas for the partition functions that follow from the trace relations are compactly summarized in terms of the Plethystic exponential
\begin{equation}
{\rm PE}\!\left[p^r Z(q^{1/r})\right]=\exp\left[\sum_{k\geq1}\frac{p^{rk}}{k}Z(q^{k/r})\right],
\end{equation}
in terms of which the partition function determined by the trace relations takes the form
\begin{equation}
\mathcal Z(p,q)=\prod_{r\geq1}{\rm PE}\!\left[p^r Z(q^{1/r})\right]\equiv \sum_{N=0}^\infty p^N Z_N(q).
\end{equation}
Indeed, expanding the exponential it is simple to recover equations \eqref{eq:Z2}, \eqref{eq:Z3} and \eqref{eq:Z4}, reproducing the partition functions found above. Taking the logarithm gives
\begin{align}
\log\mathcal Z(p,q)&=\sum_{n\geq1}p^n\left[\frac1n\sum_{r|n}r\,Z\!\left(q^{n/r^2}\right)\right].
\end{align}
Thus, defining
\begin{equation}
T_nZ(q)=\frac1n\sum_{r|n}r\,Z\!\left(q^{n/r^2}\right),
\end{equation}
For example,
\begin{equation}
T_1Z=Z(q),\qquad
T_2Z=\frac12Z(q^2)+Z(q^{1/2}),\qquad
T_3Z=\frac13Z(q^3)+Z(q^{1/3}),
\end{equation}
while
\begin{equation}
T_4Z=\frac14Z(q^4)+\frac12Z(q)+Z(q^{1/4}).
\end{equation}
Assuming this pattern to continue we recover the Hecke form of the grand partition function
\begin{equation}
\label{eq:gpf}
\mathcal Z(p,q)=\exp\left[\sum_{n\geq1}p^nT_nZ(q)\right],
\end{equation}
as in \cite{Dijkgraaf:1996xw}. 
Thus the Hecke representation itself is implied by the trace relations.

To summarize, we have shown that the above structure (of the grand canonical function) and thereby the second quantized version follows from non-linear trace relations, at finite N. See also \cite{Jevicki:1980zh} for an early discussion.

\section{BPS from Trace Relations}

We now consider a case relevant to gravity, which could be a direct precursor to a theory on $K3$ or $T^4$, with six even local channels $(0,0)\oplus 4(1,1)\oplus(2,2)$. The chiral primaries are decorated by fermionic states  $S^A$. One has two independent sets of fermionic oscillators, at each site
\begin{equation}
\{c_i^a,c_j^{b\dagger}\}=\delta^{ab}\delta_{ij},\qquad \{\bar c_i^{\bar a},\bar c_j^{\bar b\dagger}\}
=\delta^{\bar a\bar b}\delta_{ij},\qquad \{c_i^a,\bar c_j^{\bar b}\}=0, \qquad a,b,\bar a,\bar b=1,2,\label{eq:fermions}
\end{equation}
with all other anticommutators vanishing. The local operators of interest are
\begin{equation}
S_i^{(0,0)}=1,\qquad S_i^{(\bar a,b)}=\bar c_i^{\bar a}c_i^b,\qquad
S_i^{(2,2)}=\bar c_i^{\bar1}\bar c_i^{\bar2}c_i^2c_i^1.\label{eq:channels}
\end{equation}

Their multiplication closes,
\begin{equation}
S^A S^B=C_{AB}{}^C S^C,\label{eq:localmult}
\end{equation}
with
\begin{equation}
1\cdot S^A=S^A,\qquad E_{\bar a b}E_{\bar c d}=\epsilon_{\bar a\bar c}\epsilon_{bd}\,Q,\qquad
E_{\bar a b}Q=Q^2=0.\label{eq:fermionmult}
\end{equation}
The states of the model are labelled by a \emph{length} and a \emph{channel label} which runs over the six channels. To write the partition function/character we use the parameter $q$ to play the role of $e^{-\beta}$. A length-one object has the six-channel character
\begin{equation}
\chi_V(q)=1+4q^2+q^4.\label{eq:Vchar}
\end{equation}
Equivalently, the six weights are
\begin{equation}
x=(1,q^2,q^2,q^2,q^2,q^4).\label{eq:xweights}
\end{equation}
A state of length $r$ carries an additional $q^{2r-2}$, so its one-particle space is
\begin{equation}
V_r=q^{2r-2}V,\qquad\chi_{V_r}(q)=q^{2r-2}+4q^{2r}+q^{2r+2}.\label{eq:Vr}
\end{equation}
This follows from the energy assignment $\widetilde\omega_A(r)=2r+\omega_A$ with $\omega_A=(-2,0,0,0,0,2)$. The collective traces are again
\begin{equation}
\phi^{(p,\bar p)}=\sum_{h\in S_N} S(c,\bar c)\, O_{h\gamma h^{-1}}
\end{equation}
in analogy with the equation \eqref{collectivetrace}. They have been studied well enough in \cite{Jevicki:1998bm}.

Our interests are the associated the trace relations appearing at finite-$N$. The nonlinear trace relation for bosonic/fermionic impurities  are the same. They can be simplified in form as follows. Consider $\rho_r^A$ as an $r$-support collective field in channel $A$. The unrestricted collective algebra allows arbitrary products of these objects.  At finite $N$ only $N$ microscopic labels are available. The finite-$N$ evaluation map has a kernel,
\begin{equation}
\operatorname{ev}_N:{\cal A}_\infty\longrightarrow {\cal A}_{\rm micro}^{(N)},\qquad
{\cal I}_N=\ker{\rm ev}_N.\label{eq:ideal}
\end{equation}
The trace relations in the fermionic model take the same form as for the bosonic model. There are however further additional constraints that follow from the fermionic nature of the fields and there is the multiplication rule we discussed above. A product of \emph{disjoint exact supports} can survive only if the total number of labels used does not exceed $N$.  Thus,
\begin{equation}
\rho_{r_1}^{A_1}\cdots\rho_{r_s}^{A_s}=0\qquad\hbox{if}\qquad r_1+\cdots+r_s>N.\label{eq:oversupport}
\end{equation}
Also, whenever two supports overlap, their local channel labels multiply according to \eqref{eq:localmult}.  Hence overlapping products are reduced to collective fields with larger joined support.  After imposing the trace relations, the independent sectors at fixed total support $N$ are again labelled by partitions. For $m_r$ identical bosonic $r$-cycles the internal states form $\Sym^{m_r}V_r$.  Consequently the sector $\lambda$ has character
\begin{equation}
Z_\lambda(q)=\prod_{r\geq1}q^{2(r-1)m_r}\,h_{m_r}(x),\qquad x=(1,q^2,q^2,q^2,q^2,q^4),\label{eq:sectorchar}
\end{equation}
where $h_m(x)=s_{(m)}(x)$ is the complete symmetric polynomial, i.e. a Schur polynomial labelled by a Young diagram with a single row of $m$ boxes. Equation \eqref{eq:sectorchar} is the finite-$N$ counting rule obtained after the trace relations have been solved to determine a set of generating invariants.

The $h_m(x)$ are generated by
\begin{equation}
\sum_{m\geq0}h_m(x)z^m=\frac{1}{(1-z)(1-tz)^4(1-t^2z)}.
\end{equation}
The first four are (for us $t=q^2$)
\begin{align}
h_1(x)&=1+4t+t^2,\\
h_2(x)&=1+4t+11t^2+4t^3+t^4,\\
h_3(x)&=1+4t+11t^2+24t^3+11t^4+4t^5+t^6,\\
h_4(x)&=1+4t+11t^2+24t^3+46t^4+24t^5+11t^6+4t^7+t^8.\label{eq:hlist}
\end{align}

\paragraph{$N=2$:} At $N=2$, impossible support combinations are removed by trace relations. Examples of this include
\begin{equation}
\rho_1^A\rho_2^B=0,\qquad\rho_2^A\rho_2^B=0,\qquad\rho_r^A=0\quad(r>2).\label{eq:N2relations}
\end{equation}
The only surviving partitions of total support two are $[1,1]$ and $[2]$. Thus the finite-$N$ Hilbert space decomposes as
\begin{equation}
\cH_2=\Sym^2V_1\oplus V_2.\label{eq:H2}
\end{equation}
The two sector characters are
\begin{align}
Z_{[1,1]}&=h_2(x)=1+4q^2+11q^4+4q^6+q^8,\\
Z_{[2]}&=q^2\,h_1(x)=q^2+4q^4+q^6.
\end{align}
Hence
\begin{equation}
\widetilde Z_2(q)=1+5q^2+15q^4+5q^6+q^8.
\end{equation}
We can compare with $N=1$.  There is only the sector $[1]$, so
\begin{equation}
\widetilde Z_1(q)=h_1(x)=1+4q^2+q^4.
\end{equation}

\paragraph{$N=3$:} At $N=3$, the trace relations now allow a $[2,1]$ configuration but forbid, for example,
\begin{equation}
\rho_1^A\rho_3^B=0,\qquad\rho_2^A\rho_2^B=0,\qquad\rho_2^A\rho_3^B=0,\qquad
\rho_3^A\rho_3^B=0.\label{eq:N3relations}
\end{equation}
The surviving partitions are $[1,1,1]$, $[2,1]$ and $[3]$. Therefore
\begin{equation}
\cH_3=\Sym^3V_1\oplus(V_2\otimes V_1)\oplus V_3.\label{eq:H3}
\end{equation}
The three characters are
\begin{align}
Z_{[1,1,1]}&=h_3(x)=1+4q^2+11q^4+24q^6+11q^8+4q^{10}+q^{12},\\[1mm]
Z_{[2,1]}&=q^2\,h_1(x)^2=q^2+8q^4+18q^6+8q^8+q^{10},\\[1mm]
Z_{[3]}&=q^4\, h_1(x)=q^4+4q^6+q^8.
\end{align}
Adding the three finite-$N$ sectors,
\begin{equation}
\widetilde Z_3(q)=1+5q^2+20q^4+46q^6+20q^8+5q^{10}+q^{12}.
\end{equation}

\paragraph{$N=4$:} At $N=4$, two qualitatively new disjoint-support products survive: two two-cycles and a three-cycle together with a one-cycle.  The five partitions are $[1,1,1,1]$, $[2,1,1]$, $[2,2]$, $[3,1]$, $[4]$. The quotient Hilbert space is therefore
\begin{equation}
\cH_4=\Sym^4V_1\oplus\left(V_2\otimes\Sym^2V_1\right)\oplus\Sym^2V_2\oplus
(V_3\otimes V_1)\oplus V_4.\label{eq:H4}
\end{equation}
The five sector characters are
\begin{align}
Z_{[1,1,1,1]}&=h_4(x)=1+4q^2+11q^4+24q^6+46q^8+24q^{10}+11q^{12}+4q^{14}+q^{16},\\[1mm]
Z_{[2,1,1]}&=q^2\,h_2(x)h_1(x)=q^2+8q^4+28q^6+52q^8+28q^{10}+8q^{12}+q^{14},\\[1mm]
Z_{[2,2]}&=q^4 h_2(x)=q^4+4q^6+11q^8+4q^{10}+q^{12},\\[1mm]
Z_{[3,1]}&=q^4h_1(x)^2=q^4+8q^6+18q^8+8q^{10}+q^{12},\\[1mm]
Z_{[4]}&=q^6h_1(x)=q^6+4q^8+q^{10}.
\end{align}
Summing,
\begin{align}
\widetilde Z_4&=Z_{[1,1,1,1]}+Z_{[2,1,1]}+Z_{[2,2]}+Z_{[3,1]}+Z_{[4]}\nonumber\\
&=1+5q^2+21q^4+65q^6+131q^8+65q^{10}+21q^{12}+5q^{14}+q^{16}.
\end{align}
$[2,2]$ and $[3,1]$ are the new disjoint-support structures which first become possible at $N=4$.  In the collective description they are independent finite-$N$ sectors.

The pattern is clear.  Once the trace relations have reduced the Hilbert space to disjoint exact-support sectors, we obtain
\begin{equation}
\cH_N=\bigoplus_{\lambda=(1^{m_1}2^{m_2}\cdots)\vdash N}\bigotimes_{r\geq1}\Sym^{m_r}V_r.\label{eq:generalH}
\end{equation}
Taking the graded character gives
\begin{equation}
\widetilde Z_N(q)=\sum_{\substack{m_1,m_2,\ldots\geq0\\ \sum_{r\geq1}r\,m_r=N}}\prod_{r\geq1}
q^{2(r-1)m_r}h_{m_r}(1,q^2,q^2,q^2,q^2,q^4).\label{eq:generalZN}
\end{equation}
This is equivalent to
\begin{equation}
\widetilde Z_N(q)=q^{2N}\sum_{\substack{m_1,m_2,\ldots\geq0\\ \sum r\,m_r=N}}\prod_{r\geq1}
s_{(m_r)}(q^{-2},1,1,1,1,q^2).\label{eq:desired}
\end{equation}
Thus the Schur-polynomial formula has a direct finite-$N$ trace-relation interpretation: $m_r$ is the number of surviving disjoint exact supports of size $r$, and the complete symmetric polynomial $h_{m_r}=s_{(m_r)}$ counts the bosonic symmetrization of their six internal channel labels. Finally, summing the equation \eqref{eq:generalZN} over $N$ gives
\begin{align}
\sum_{N\geq0}p^N\widetilde Z_N(q)&=\prod_{r\geq1}\left[\sum_{m\geq0} h_m(x)\left(p^r q^{2r-2}\right)^m\right]=\prod_{r\geq1}\prod_{a=1}^6\frac{1}{1-p^r q^{2r-2}x_a}\nonumber\\
&=\prod_{r\geq1}\frac{1}{(1-p^rq^{2r-2})(1-p^rq^{2r})^4(1-p^rq^{2r+2})},
\end{align}
which is the correct six channel truncation of the grand partition function of \cite{deBoer:1998us}. 

\section{Thermodynamics}

The preceding sections show that the non-linear finite-$N$ trace relations lead to the partition-labelled normal forms and hence to the grand-canonical/Hecke partition function.  We now use that result to study its thermodynamic consequences, in particular the presence of phase transition~\cite{Shenker:2011zf}. For general orbifold CFT's this has been seen in~\cite{Keller:2011xi}. That it indeed happens even for the present, toy model (consisting mainly of trace relations) is a further verification of their relevance.

\subsection{Finite-$N$ stability and the large-$N$ limit}

Consider a two-dimensional bosonic oscillator seed theory,
\begin{equation}
\label{eq:seed}
Z(q)=\sum_{m=0}^{\infty}d(m)q^{\omega(m)}=\frac{1}{q(1-q)^2},
\qquad d(m)=m+1,\qquad \omega(m)=m-1.
\end{equation}
Substituting this seed into \eqref{eq:gpf} gives
\begin{equation}
\mathcal Z(p,q)=\prod_{m=0}^{\infty}\prod_{n=1}^{\infty}
\left(1-p^nq^{\omega(m)/n}\right)^{-d(m)}.
\end{equation}
It is convenient to remove the vacuum energy by defining
\begin{equation}
\label{eq:sgpf}
\widetilde{\mathcal{Z}}(p,q)=\mathcal Z(pq,q)
=\prod_{m=0}^{\infty}\prod_{n=1}^{\infty}
\left(1-p^nq^{\widetilde\omega(m,n)}\right)^{-d(m)},
\end{equation}
where
\begin{equation}
\label{eq:spectrum}
\widetilde\omega(m,n)=\frac{\omega(m)}{n}+n.
\end{equation}
Writing $p=e^{-\mu}$ and $q=e^{-\beta}$, the fixed-$N$ partition function is extracted by
\begin{equation}
\label{eq:residue}
\widetilde Z_N(q)=\frac{1}{2\pi i}\oint\frac{dp}{p^{N+1}}\widetilde{\mathcal{Z}}(p,q).
\end{equation}
Its low-temperature expansion stabilizes with $N$:
\begin{equation}
\widetilde Z_N(q)=\widetilde Z_{N-1}(q)
+\mathcal O\!\left(q^{N-\frac12\lfloor N/2\rfloor}\right).
\label{eq:stability}
\end{equation}
For example,
\begin{align}
\widetilde Z_3(q)&=1+2q+q^{3/2}+8q^2+5q^{5/2}+q^{8/3}+24q^3
+\mathcal O(q^{3+\epsilon}),\\
\widetilde Z_4(q)&=1+2q+q^{3/2}+8q^2+5q^{5/2}+q^{8/3}+25q^3
+\mathcal O(q^{3+\epsilon}).
\end{align}
The agreement through the predicted stable range is explicit. On the other hand the infinite -$N$ partition function reads
\begin{equation}
\widetilde Z_\infty(q)=\prod_{m=0}^{\infty}\prod_{n=1}^{\infty}
\left(1-q^{\widetilde\omega(m,n)}\right)^{-d(m)},
\end{equation}
representing the large $N$ limit.

\subsection{Bose--Einstein condensation and the double-scaling limit}

The thermodynamic transition follows directly from the grand partition function whose structure was fixed above by the finite-$N$ trace relations.  The shifted spectrum \eqref{eq:spectrum} has a unique zero mode at $(m,n)=(0,1)$. Separation of this reservoir from the positive-energy excitations gives
\begin{equation}
\widetilde{\mathcal{Z}}(p,q)=\widetilde{\mathcal{Z}}_{\rm gd}(p)\widetilde{\mathcal{Z}}_{\rm ex}(p,q),
\end{equation}
with
\begin{equation}
\widetilde{\mathcal{Z}}_{\rm gd}(p)=\frac{1}{1-p},\qquad
\widetilde{\mathcal{Z}}_{\rm ex}(p,q)=\prod_{(m,n)\neq(0,1)}
\left(1-p^nq^{\widetilde\omega(m,n)}\right)^{-d(m)}.
\end{equation}
The charge stored in the excited sector is
\begin{equation}
N_{\rm ex}(p,q)=p\frac{\partial}{\partial p}\log\widetilde{\mathcal{Z}}_{\rm ex}(p,q)
=\sum_{(m,n)\neq(0,1)}
\frac{n(m+1)}{p^{-n}q^{-\widetilde\omega(m,n)}-1}.
\end{equation}
It reaches its maximal capacity at $p\to1^{-}$,
\begin{equation}
N_{\rm ex}^{\rm max}(\beta)=
\sum_{(m,n)\neq(0,1)}
\frac{n(m+1)}{e^{\beta\widetilde\omega(m,n)}-1}.
\end{equation}
For $N<N_{\rm ex}^{\rm max}$ there is an interior saddle $p^*<1$.  For
$N\geq N_{\rm ex}^{\rm max}$ the saddle reaches $p^*=1^{-}$ and the excess charge occupies the zero-mode reservoir,
\begin{equation}
\label{eq:saddle}
N_{\rm gd}(\beta)=N-N_{\rm ex}^{\rm max}(\beta).
\end{equation}
At high temperature,
\begin{equation}
\begin{aligned}
N_{\rm ex}^{\rm max}(\beta)
&\simeq \sum_{l=1}^{\infty}\sum_{n=1}^{\infty}ne^{-l\beta n}
\sum_{m=0}^{\infty}(m+1)e^{-l\beta(m-1)/n}\\
&\simeq \sum_{l=1}^{\infty}\frac{6}{(l\beta)^6}
=\frac{6\zeta(6)}{\beta^6}.
\end{aligned}
\end{equation}
Thus
\begin{equation}
\beta_c=\lambda_c N^{-1/6},\qquad
\lambda_c=(6\zeta(6))^{1/6}.
\end{equation}
In the double-scaling limit
\begin{equation}
\beta\to0^+,\qquad N\to\infty,\qquad \lambda=N^{1/6}\beta\ \text{fixed},
\end{equation}
the condensate fraction becomes
\begin{equation}
\label{eq:fraction}
\frac{N_{\rm gd}(\lambda)}{N}
=\max\!\left(1-\left(\frac{\lambda_c}{\lambda}\right)^6,0\right).
\end{equation}
The same result follows from the saddle-point evaluation of equation \eqref{eq:residue}; the saddle moves to the boundary $p=1$ at $\lambda=\lambda_c$.

\subsection{Macroscopic cycles and the confinement/deconfinement order parameter}

The condensation also has a direct description in terms of permutation cycles.  Introduce
\begin{equation}
\label{eq:order}
G_\infty(\lambda)=\lim_{\epsilon\to0^+}\lim_{N\to\infty}G_N^\epsilon(\lambda),
\qquad
G_N^\epsilon(\lambda)=\frac1N\sum_{M\geq\epsilon N}^{N}
M\langle n_M(\lambda)\rangle_N,
\end{equation}
which measures the fraction of replicas belonging to macroscopic cycles.  The order of the two limits is important.  From cycle decomposition, the expectation value of the number of $M$-cycles is
\begin{equation}
\label{eq:number}
\langle n_M(\lambda)\rangle_N
=\frac{T_M \widetilde Z(\lambda)\widetilde Z_{N-M}(\lambda)}
{M\widetilde Z_N(\lambda)},
\end{equation}
where the tildes indicate the quantities constructed from the shifted spectrum.
Separating zero-mode and excited cycles,
\begin{equation}
n_M=n_M^{\rm gd}+n_M^{\rm ex},
\end{equation}
one finds that the macroscopic excited-cycle contribution is suppressed, while
\begin{equation}
\langle n_M^{\rm gd}(\lambda)\rangle_N
=\frac1M\langle 1_{N_{\rm gd}\geq M}\rangle_N.
\end{equation}
Consequently the macroscopic-cycle order parameter is precisely the condensate fraction,
\begin{equation}
G_N^\epsilon(\lambda)=
\max\!\left(\frac{N_{\rm gd}(\lambda)}{N}-\epsilon,0\right),
\qquad
G_\infty(\lambda)=\frac{N_{\rm gd}(\lambda)}{N}.
\end{equation}
Thus the onset of the condensate is simultaneously the appearance of macroscopic permutation cycles.  In the orbifold interpretation this is the natural confinement/deconfinement order parameter: below the transition the ensemble is a gas of microscopic cycles, while above it a finite fraction of the charge is carried by macroscopic cycles.

\begin{figure}[t]
\centering
\includegraphics[width=0.72\linewidth]{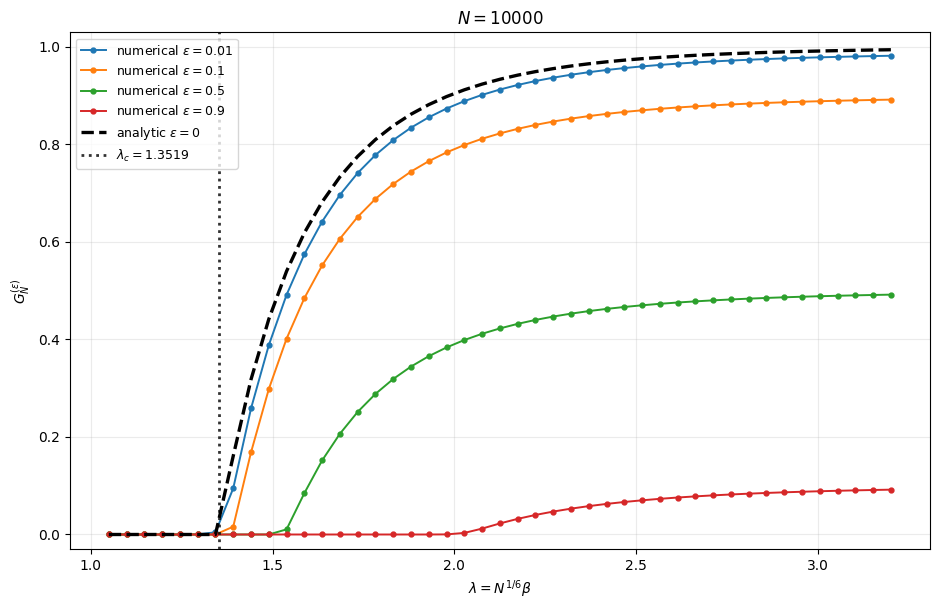}
\caption{Macroscopic-cycle order parameter across the transition.  The finite-$N$ numerical result approaches the analytic condensate fraction in the double-scaling limit.}
\label{fig:transition}
\end{figure}

Using equation \eqref{eq:number} to evaluate equation \eqref{eq:order} numerically gives the finite-$N$ curves shown in figure \ref{fig:transition}, which approach the analytic result \eqref{eq:fraction}. This behaviour is similar to other cases with Hironaka structure \cite{Ahn:2026ohj}.

\acknowledgments
RdMK and AJ would like to thank Kyung Hee University and KIAS for their hospitality and support, during which this work was completed. The work of RdMK was supported by a start-up research fund of Huzhou Normal University, a Zhejiang Province talent award and a Changjiang Scholar award.

\end{document}